\documentstyle{article}

\title{Entropy is complexity}
\author{Adonai S. Sant'Anna\thanks{Permanent address:
Department of Mathematics, Federal University of Paran\'a, P. O.
Box 019081, Curitiba, PR, 81531-990. E-mail: adonai@ufpr.br.}}

\date{Department of Philosophy\\University of South Carolina\\Columbia, SC, 29208}
\begin{document}

\newtheorem{definicao}{Definition}
\newtheorem{teorema}{Theorem}
\newtheorem{lema}{Lemma}
\newtheorem{corolario}{Corolary}
\newtheorem{proposicao}{Proposition}
\newtheorem{axioma}{Axiom}
\newtheorem{observacao}{Observation}

\maketitle


\begin{abstract}

In a recent paper Andrei N. Soklakov explained the foundations of
the Lagrangian formulation of classical particle mechanics by
means of Kolmogorov complexity. In the present paper we use some
of Soklakov ideas in order to derive the second law of
thermodynamics. Our main result is that the complexity of a
thermal system corresponds to its entropy.

\end{abstract}

\section{Introduction}

By means of the so-called prefix version of Kolmogorov complexity,
introduced by Levin \cite{Levin-74,Levin-76}, G\'acs
\cite{Gacs-74}, and Chaitin \cite{Chaitin-69}, Soklakov
\cite{Soklakov-02} was able to explain why the Lagrangian $L$ of a
composite system always has the form $L = L_1+L_2-V$, where $L_1$
and $L_2$ are the Lagrangians of free subsystems and $V$ accounts
for the interaction part.

One of the key aspects of Soklakov's work is that complexity is
physically interpreted as energy. In this paper we extend his
ideas in order to ground the concept of entropy by means of the
notion of complexity.

\section{Mathematical background}

This section is a brief review of some parts of
\cite{Soklakov-02}.

Let $X$ be the set of all finite binary strings $\{\Lambda$, 0, 1,
00, 01, 10, 111, 000, 001, ...$\}$, where $\Lambda$ is the string
of length zero. Let $Y$ be a subset of $X$ such that no string in
$Y$ is a prefix of another. From now on we will consider only
prefix computers, i.e., partial recursive functions $C:Y\times
X\to X$. This is a very weak restriction from the theoretical
point of view. $C(p,d) = \alpha$ means that $\alpha$ is the output
of the computation of the data string $d$ with the program string
$p$ by means of the computer $C$. The complexity of $\alpha$,
given $d$, and relative to the computer $C$, is given by:

\begin{equation}
K_C(\alpha|d) = \mbox{min}\{ |p| \mbox{ such that } C(p,d) =
\alpha\},
\end{equation}

\noindent where $|p|$ denotes the length of the program $p$ in
bits. It is well known that there is an optimal computer $U$ for
which $K_U(\alpha|d)\leq K_C(\alpha|d) + \kappa$, where $\kappa$
is a constant that depends on $C$ and $U$, only. Any prefix
computer can be simulated by $U$, and $U$ is called a universal
prefix computer.

In order to adopt a simpler notation, we make $K(\alpha|d) =
K_C(\alpha|d)$.

One possible intuitive meaning for $K(\alpha|d)$ is that it
corresponds to the big picture of a very detailed object $\alpha$
with a previous past given by $d$. It is worth to remark that the
term ``past'' does not entail any corresponding notion of time.
The point is that there is some kind of causation between $d$ and
$\alpha$. In \cite{Soklakov-02} Soklakov describes this causation
by means of time. But in this paper we use another parameter (with
another interpretation) to relate $d$ and $\alpha$.

Let $^Jg:X\to X_1\times ... \times X_J (X_j = X)$ be a function.
The complexity of $^Jg$ at $x_0\in X$ is defined as

\begin{equation}
K_{x_0}[^Jg] = \frac{1}{J}\sum_{k=0}^{J-1}K(x_{k+1}|x_k),
\end{equation}

\noindent where $\{ x_1, x_2, ..., x_J\} =$ $^Jg(x_0)$ and
$K(x_{k+1}|x_k)$ is the complexity of $x_{k+1}$ given data $x_k$,
with respect to a universal prefix computer.

This last equation will be very important in our derivation of the
second law of thermodynamics.

\section{Entropy}

In this section we intend to use complexity theory in
thermodynamics. The first important problem is the physical
interpretation of the involved mathematical concepts. Since we are
interested on thermal systems, we interpret the strings of $X$ as
possible microstates of a given thermal system. So, let

\begin{equation}\label{equacao1}
K_{x_0}^{t_i}[^Tg] =
\frac{1}{T}\sum_{k=0}^{T-1}K^{t_i}(x_{k+1}|x_k)
\end{equation}

\noindent denote the complexity of a thermal system at absolute
zero (the temperature is ideally zero Kelvin), with respect to the
instant of time $t_i$ ($i$ stands for initial).
$K^{t_i}(x_{k+1}|x_k)$ is the complexity of the microstate
$x_{k+1}$ given the microstate $x_k$, at the initial instant
$t_i$. It is important to remark that function $^Tg$ corresponds
to a dynamical evolution which starts at zero Kelvin. This means
that we are adopting temperature as a parameter for describing
such a dynamics. In \cite{Soklakov-02} Soklakov used time as the
parameter for describing the dynamics of a mechanical system. The
fact that the summation starts at zero ($k = 0$) just reflects the
absolute nature of temperature. So, $T$ denotes an absolute value
of temperature, although we are not talking about the Kelvin
measurement scale. This absolute scale of temperature is zero when
the temperature is zero Kelvin. But the main difference is that
$T$ is measured in discrete quantities, since we are talking about
a summation.

Now let

\begin{equation}\label{equacao2}
K_{x_0}^{t_f}[^Tg] =
\frac{1}{T}\sum_{k=0}^{T-1}K^{t_f}(x_{k+1}|x_k)
\end{equation}

\noindent denote the complexity of a thermal system at absolute
zero, with respect to the instant of time $t_f$ ($f$ stands for
final). Again we insist that the term ``complexity of a thermal
system at absolute zero'' does not mean that the thermal system
has a zero Kelvin temperature at instant $t_f$. It just means that
the dynamics of the thermal system started at the zero point.

According to equations (\ref{equacao1}) and (\ref{equacao2}) we
have:

\begin{equation}\label{equacao3}
K_{x_0}^{t_f}[^Tg] - K_{x_0}^{t_i}[^Tg] =
\frac{1}{T}\sum_{k=0}^{T-1}\left(K^{t_f}(x_{k+1}|x_k) -
K^{t_i}(x_{k+1}|x_k)\right).
\end{equation}

The physical interpretation of this equation seems to be quite
natural. If the microstates $x_k$ ($k = 0, 1, 2, ..., T-1$) give
us the detailed description of the thermal system, then
$K_{x_0}^{t_f}[^Tg]$ and $K_{x_0}^{t_i}[^Tg]$ give the big picture
of the same system at different instants of time. This big picture
is the macrostate of the thermal system, i.e., the resulting
energy associated to the system. So, we interpret the left side as
entropy, and the difference between summations on the right side
corresponds to the heat absorbed by the thermal system during the
time interval $(t_i,t_f)$. Instead of a statistical or
probabilistic approach to thermodynamics, this suggests a
computational approach with analogous results.

\section{Remarks}

We make here some final remarks:

\begin{enumerate}

\item In \cite{Soklakov-02} the author uses a very important
physical principle which he calls the {\em simplicity principle\/}
(SP). According to SP ``among all dynamical laws that are
consistent with all the other axioms [of the theory], the laws
with the smallest descriptional complexity predominate the
system's behavior''. Actually this is a version of Occam's razor
principle, according to which simple theories are more economical
and are usually better suited for making predictions. Soklakov
uses SP in order to justify why it is necessary to minimize a
functional of action which is an integral of the Lagrangian of the
mechanical system that he studies. In our paper we make no use of
SP. This is so because we are concerned with a difference of
complexities and not with any complexity itself. Thus, any process
of minimization seems to be unnecessary here.

\item Many authors make a relationship between entropy and the
direction of time. In equation (\ref{equacao3}) this relationship
is made explicitly.

\item Equation (\ref{equacao3}) is an explicit formula that
relates microstate ($x_k$) and macrostate (temperature and heat).
Nevertheless, like temperature, heat (energy) is discrete,
although equation (\ref{equacao3}) is valid for any unit of
measurement for energy if the unit measurement for temperature
remains absolute, i.e., it is zero at the zero absolute
temperature.

\item There is no need of the use of differential and integral
calculus in our approach. In \cite{Soklakov-02} the author
understands the necessity for a formulation of the Galilean
relativity principle in a discrete form, since the main basis of
complexity theory is discrete mathematics. According to him, this
problem will be considered in future research.

\item Our main result is that entropy may be understood as the
complexity of a thermal system.

\end{enumerate}

\section{Acknowledgements}

This work was partially supported by CAPES (Brazilian government
agency).

\end{document}